# Measures for Classification and Detection in Steganalysis

Sujit Gujar and C E Veni Madhavan

Dept of Computer Science and Automation
Indian Institute of Science, Bangalore,560012.
{sujit,cevm}@csa.iisc.ernet.in

**Abstract.** ***Still and multi-media images are subject to transformations for compression, steganographic embedding and digital watermarking. In a major program of activities we are engaged in the modeling, design and analysis of digital content. Statistical and pattern classification techniques should be combined with understanding of run length, transform coding techniques, and also encryption techniques.***

## 1 Introduction

*Steganography* is the art and science of secret communication, aiming to conceal the existence of the communication. *Steganalysis* is the art of seeing unseen. With advent of computers, hiding information inside digital carriers, especially multi media files like audio (*.wav*) files, images(*.bmp, .pnm, .jpg*), is becoming popular( [1,5]). Digital images are most common sources for hiding message. The process of hiding information is called embedding. Least Significant Bit (LSB) embedding is the most widely used steganographic technique. In LSB embedding, the LSB of uncompressed images are replaced with the message bits. The amount of embedding (the number of bits embedded) referred to as *level*, is given as a percentage of the total number of pixels.
Some of the powerful methods for the analysis of steganographic images are [6], [7], [2]. We propose new measures and techniques for detection and analysis of steganographic embedded content. We show that both statistical and pattern classification techniques using our proposed measures provide reasonable discrimination schemes for detecting embeddings of different levels. Our measures are based on a few statistical properties of bit strings and wavelet coefficients of image pixels.
In Section 2, we explain our approach towards classification of given data based on a feature vector consisting of statistical measures and using Support Vector Machine (SVM) tools. In Section 3 we propose the use of wavelet transforms for steganalysis. Our results presented in Section 2 and 3 show the efficacy of our measures in discriminating different levels of embedding. We conclude with our plans for improved and finer steganalysis in section 4

This paper is extension of our paper [4]. In this paper we provide analysis of proposed measure based on wavelets in [4]. For more detailed report, readers are referred to [3]

## 2 Classification Of The Given Data Using Statistics And SVM

### 2.1 Classification Of Different Files

Steganography is one kind of transformation of one data into other with the help of a cover. So, if messages before embedding can be classified, we believe that we can classify them in transformed domain as well. Now assume, we can classify the data in transformed domain into different classes. What next? Why do we need classification in first place? Classification is required because, if we classify given part of data, we can narrow down the possible transformed space and we can predict what to except next and what should not be next. This will help us to predict message bits.

Can we really classify data? We use a statistical feature space. We propose a vector of statistical measures [6] for this purpose. Our feature vector consists of nine statistical measures. Thus, $\mu \in \mathbb{R}^9$. The measures are as follows.

$\mu_1$ : *Weighted sum of the range of k-gram frequencies.* Let $f(k, j)$ denote the overlapping frequency of the k-gram binary pattern of the integer $j$ in $S_i$. For example $f(4, 3)$ = Number of occurrences of the patterns $< 0011 >$ in $S_i$. For a 32 bit word W, we define

$$\mu_1(W) = \sum_{k=1}^{4} (\max_j(f(k, j)) - \min_j(f(k, j))) 2^{4(k+1)}$$

We expect the measure $\mu_1$ to be smaller for random strings as compared to non-random strings.

$\mu_2$ : *Weighted sum of run lengths.* Let the vector $< l_1, l_2, ... >$ denote the sequence of run lengths of $0'$s and $1'$s in a word W. Then we define,

$$\mu_2(W) = \sum 2^{c_i l_i}$$

where $c_i$ are specifically chosen weights. We set $c_i = 1 \; \forall \; i$, without loss of generality. For random strings, we expect the measure $\mu_2$ to be smaller compared to non-random strings, since one expects very few long runs.

$\mu_3$ : *Weighted sum of byte-wise hamming weight transition.* Let W $= < b_0, b_1, b_2, b_3 >$, where $b_i$'s are the bytes of the 32 bit word. Let $\#1(b)$ denote the number of 1's in a 8 bit byte. Then we define,

$$\mu_3(W) = 2^{\#1(b_0)} + 2^{\#1(b_0 \oplus b_1)} + 2^{\#1(b_1 \oplus b_2)} + 2^{\#1(b_2 \oplus b_3)}$$

For random strings W, we expect $\mu_3(W)$ to be higher than non-random strings. It is also possible to define the measure $\mu_3$ with respect to overlapping bytes in a word, to measure the smoothness/suddenness of transitions.

$\mu_4$ : *Fourier transform of the autocorrelation function of the sequence bits* in W. Let W $= < a_0, ..., a_{31} >$ be a 32 bit word. The autocorrelation function $A(W)$ is the sequence $A(W) =< c_0, .., c_{31} >$ where $c_i = \sum_{j=0}^{31} a_j.a_{j+1} \pmod{32}$, $i = 0, .., 31$ and the multiplication operation is over $F_2$ vectors. The discrete Fourier transform $F(A(W))$ is given by the sequence $F(A(W)) =< f_0, ..., f_{31} >$; where $f_k = \sum_{j=0}^{31} c_j \, \omega^{jk \, mod 32}$ $k = 0, ..., 31$. Here $\omega$ is a $32^{nd}$ root of unity. Finally, the measure $\mu_4$(W) is a root mean square average of $F$ and is given by,

$$\mu_4(W) = (\sum_{j=0}^{31} |f_j|^2)^{1/2}$$

For random string W, we expect $\mu_4(W)$ to be smaller than non-random strings.

$\mu_5$ : *Weighted Hadamard transform.* Using an 8x8 Hadamard matrix $(H)$ and operation $y = Hx$, where $x$ is 8x1 bit vector, we get measure $\mu_5$. $x$ is single data byte. Especially when,

Hadamard transform is applied on images, $x$ is pixel value.

$\mu_6$, $\mu_7$, $\mu_8$, $\mu_9$ : These measures are based on weighted entropy measure $\sum p_i \log p_i$ where $p_i$'s are probabilities of 1,2,3,4 grams. The weights are chosen experimentally to amplify the range.

SVM (***S**upport **V**ector **M**achine*) is a powerful tool for pattern classification. With introduction of kernel tricks in SVM, it has become a very popular in machine learning community. In some cases, the given data is not directly classifiable. Such cases can be solved by transforming the given data to higher dimensional space in such a way that in transformed domain, the classification is much easier. Kernel tricks help this without actually transforming features to higher dimensional space. The above statistics i.e. $\mu$ is used as feature vector of the data. For training of SVMs, we measure statistics on 2000 words (8000bytes) of 30 different files to get 30 different $\mu$ and same for each class. For testing, we measure statistics on 2000 words of 20 different files from each class. Though we have used measures calculated on 2000 words, experiments shows that 400 words are sufficient for testing a data for classification. SVM tool is taken from *http://www.csie.ntu.edu.tw/~cjlin/libsvm/*. We are using the most widely used 'Gaussian kernel' for SVM. For avoiding some features dominating in classification, we scale $\mu$ to zero mean, unit variance. We use the following 8 different classes:

1. jpeg 2. bmp/pnm 3. zip files 4. gz files 5. text files 6. ps files 7. pdf files and 8. c files.

We present the result in confusion matrix format. $ij^{th}$ entry is probability of data belonging to class $i$ and getting classified as class $j$. Table 1 shows the result. It can be seen that as needed, $ii^{th}$ entry is very near to 1 for most of the classes.

**Table 1.** Confusion Matrix For Data Classification

| | | | | | | | |
|---|---|---|---|---|---|---|---|
| 0.9 | 0.05 | 0.05 | 0.0 | 0.0 | 0.0 | 0.0 | 0.0 |
| 0.0 | 0.9 | 0.0 | 0.0 | 0.0 | 0.0 | 0.05 | 0.05 |
| 0.0 | 0.0 | 0.6 | 0.35 | 0.0 | 0.0 | 0.05 | 0.0 |
| 0.0 | 0.0 | 0.1 | 0.9 | 0.0 | 0.0 | 0.0 | 0.0 |
| 0.0 | 0.0 | 0.0 | 0.0 | 1.0 | 0.0 | 0.0 | 0.0 |
| 0.0 | 0.0 | 0.0 | 0.0 | 0.05 | 0.95 | 0.0 | 0.0 |
| 0.0 | 0.0 | 0.6 | 0.05 | 0.0 | 0.05 | 0.3 | 0.0 |
| 0.0 | 0.0 | 0.0 | 0.0 | 0.05 | 0.0 | 0.0 | 0.95 |

We used a total of 180 files for testing and were able classify with 82.22% accuracy.

### 2.2 Analysis of LSB Planes From Stegoed and non-Stegoed Images

In above experiments, we measure statistics on whole sequence of bits of the given data. An embedding operation is performed on LSBs of an image. So to capture perturbation due to steganographic operation, we measure statistics only of LSBs of images in pursuit of our aim, detecting levels of embedding in an given image. In this direction, we first considered only two classes of LSBs, one is LSB planes obtained from non-stegoed image and other LSB planes obtained from images with 50% embedding. The same $\mu$ defined above is measured on LSB planes of 30 images from both classes. (total 180 = 30*3(colors/images)*2 classes). Out of these, 150 are used for training SVM and testing was performed on 30. Here, we have two classes

1. LSB plane of non-Stegoed image. 2. LSB plane of stegoed image.

We present the results in confusion matrix in Table 2

**Table 2.** Confusion Matrix For 2 Category LSB Classification

| | |
|---|---|
| 0.67 | 0.33 |
| 0.0 | 1.0 |

Overall accuracy is 85%. Being motivated by the results, we now consider this as 4 category classification problem. The different classes defined are,

1. LSB plane of non-Stegoed image. 2. LSB plane of 25% stegoed image. 3. LSB plane of 50% Stegoed image. 4. LSB plane of 75% stegoed image.

The confusion matrix for this experiment is in Table 3,

**Table 3.** Confusion Matrix For 4 Category LSB Classification

| | | | |
|---|---|---|---|
| 0.6 | 0.0 | 0.33 | 0.07 |
| 0.0 | 0.6 | 0.27 | 0.13 |
| 0.0 | 0.0 | 0.4 | 0.6 |
| 0.0 | 0.0 | 0.0 | 1.0 |

The overall efficiency is 65%. Thus, this experiment alone is not sufficient for detection of levels of embedding. Hence we take another alternative approach.

## 3 Analysis Of Images Using Wavelet Transforms

Our feature vector $\mu$ considers a linear sequence of bits as input. However, image properties are in general captured more accurately by two dimensional transforms. Our goal is to classify images accurately under different levels of embedding. The approaches in Section 2.1 and 2.2 serve as good handles in this direction.

To further enhance our understanding of the effects of embedding, we study the behavior of wavelet coefficients. Farid et al [5,8] have shown that wavelet domain can capture image characteristics, such as whether an image is a natural image or a computer generated one or is a scanned one. They have shown that the feature vector given by them can be used for universal steganalysis. Their aim was only to find whether an image contains any kind of hidden information or not. We further explore the detection of the level of embedding.

### 3.1 Hypothesis

Our motivation to study the wavelet domain, rather than pixels directly, is that the averages in wavelet coefficients smoothen the pixel values and hence *it is expected* that even minor anomalies in neighboring pixels introduced by stego operation would lead to *amplified* changes in the wavelet domain. We intend to capture and attempt to calibrate these changes w.r.t graded embedding. We consider second level wavelet sub-bands of images. The *Haar wavelet* is used as the mother wavelet.

### 3.2 Notations

For our experiments, we use 15 images that do not contain any hidden information. These are images taken with a Nikon Coolpix camera at full 8M resolution with most of images stored in RAW format. These images are then cropped to get 800x600 images without doing any image processing operations. Let,

$$\begin{aligned}
\mathrm{I} = \{I_j : j = 0, 1, 2, \ldots, 14\} &\text{ be the set of natural unstegoed images.}\\
k &: \text{The initial LSB embedding present in an given image. i.e. } k\% \text{ LSB's}\\
&\quad \text{of an image have been modified by steganographic operations.}\\
S_k &: \text{The Start Image, that is an image } \in \mathrm{I} \text{ with } k\% \text{ embedding.}\\
&\quad (k \text{ is the unknown to be detected.})\\
i &: \text{The forced embedding level.}\\
&\quad \text{(will be defined in Section 3.3)}\\
S_{ki} &: \text{An image } \in \mathrm{I} \text{ with } k\% \text{ original embedding and } i\% \text{ forced embedding.}
\end{aligned}$$

### 3.3 Our Approach

Let $S_k$ be the given image. We call this as the start image. We do additional embedding on it to get $S_{ki}$ and refer this kind of embedding as '*forced embedding*'. Our approach is to compute some transforms on both $S_k$ and $S_{ki}$, and study a measure of the difference between the transform coefficients for finding $k$. This procedure is explained with the help of Fig. 1. In Fig. 1 the transform used is the wavelet transform.

### 3.4 Definitions

We consider second level wavelet sub-bands. So, each $4 * 4$ block in images will contribute to exactly one wavelet coefficient in each sub-band viz. LL, LH, HL, HH. Let, We consider the $2^{nd}$ level LL sub-band coefficients, since most of the energy gets concentrated in this sub-band. Let a $4 * 4$ subblock of an image be denoted by :

$$\begin{pmatrix} a & b & c & d \\ e & f & g & h \\ i & j & k & l \\ m & n & o & p \end{pmatrix}$$

The $2^{nd}$ level LL wavelet coefficient is given by

$$\frac{1}{4} * (a + b + c + d + e + f + g + h$$
$$+ i + j + k + l + m + n + o + p)$$

(Note : The $2^{nd}$ Level LL sub-band size is $\frac{1}{4}^{th}$ of the original image size in both directions.)

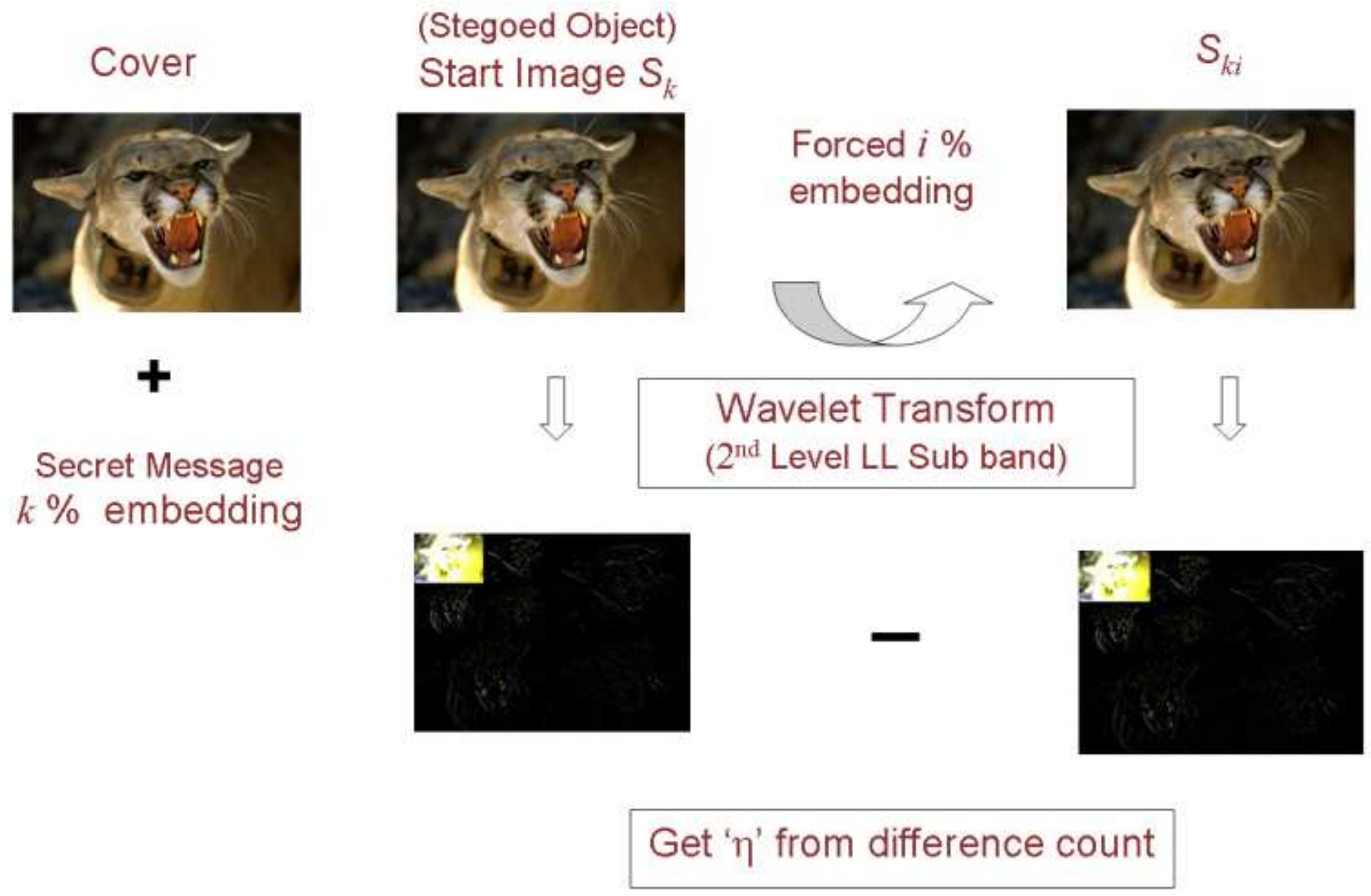

**Fig. 1.** *The process to get* $\eta$

The LH coefficient is given by

$$\frac{1}{4} * \{(a+b+e+f++i+j+m+n) - (c+d+g+h+k+l+o+p)\}$$

The HL coefficient is given by

$$\begin{gathered}\tfrac{1}{4} * \{(a+b+c+d+e+f+g+h) \\ -(i+j+k+l+m+n+o+p)\}\end{gathered}$$

The HH coefficient is given by

$$\begin{gathered}\tfrac{1}{4} * \{(a+b+e+f+k+l+o+p) - \\ (c+d+g+h+i+j+m+n)\}\end{gathered}$$

Let the image $S_k$ be considered as made up of $4*4$ blocks.

$$\begin{gathered}\text{Let } P \text{ denote a } 4*4 \text{ block in } S_k.\ P = (u_{ij}) \\ \text{and } P' \text{ denote corrosponding } 4*4 \text{ block in } S_{ki}.\ P' = (u'_{ij})\end{gathered}$$

We define the following random variables,

$$X_0 = \#\{\mid \sum \left(u_{ij} - u'_{ij}\right) \mid \neq \ 0 : \text{for all non-overlapping blocks P in } S_k\}$$
$$X_1 = \mid \sum \left(u_{ij} - u'_{ij}\right) \mid \ \text{ over non-overlapping blocks P in } S_k$$
$$X_2 = \sum \left(u_{ij} - u'_{ij}\right) \text{ over non-overlapping blocks P in } S_k$$
$$\eta = \frac{X_0 * 500}{\text{image size in pixels}}$$
$$\Gamma_W^{ki} = \text{SNR between } 2^{nd} \text{ level LL sub-band of } S_{ki} \text{ and } S_k$$

We have chosen the factor 500 to normalize the quantity $\eta$ to be near 100 for the size of images being considered $(800 * 600)$.

### 3.5 Analysis

Let,

$$\begin{aligned}
p &= \text{probability of LSB of pixel in a } 4*4\text{block be even i.e. `0' in Cover } S \\
p' &= \text{probability of LSB of pixel in the } 4*4\text{block be even i.e. `0' in } S_k \\
&= \frac{k}{2} + (1-k)*p \\
p'' &= \text{probability of LSB of pixel in the } 4*4\text{block be even i.e. `0' in } S_{ki} \\
&= \frac{i}{2} + (1-i)*p' \\
Pr &= \text{probability of a particular } 2^{nd} \text{ level LL wavelet coefficient in } S_{ki} \text{ is different} \\
&\quad \text{from corresponding wavelet coefficient in } S_k \qquad (1) \\
\text{X}_0 &= \frac{\text{Image Size in Pixels}}{4*4} * Pr \\
&\Rightarrow \\
\text{X}_0 &\propto Pr
\end{aligned}$$

Let,

$$\eta_{ki} = \eta \text{ with } k\% \text{ initial embedding and } i\% \text{ forced embedding.}$$

*Theorem* :
i. $\eta_{ki}$ increases with $i$ and decreases slightly with $k$.
ii. $\Gamma_W^{ki}$ increase with increase in $k$.

Proof: Observe that,

$$\eta_{ki} \propto \text{X}_0 \propto Pr$$

$$
\begin{aligned}
Pr = 1 - \text{prob}\{\mid \textstyle\sum (u_{ij} - u'_{ij}) \mid = \ 0\} \\
= 1 - \text{prob}\{&\text{No pixel in the particular } 4*4 \\
&\text{block has been replaced with data bits} \\
&\ \text{OR} \\
&\text{2 pixels have been replaced with data bits} \\
&\text{in such way that one pixel value} \\
&\text{increases by 1 and other decreases by 1} \\
&\ \text{OR} \\
&\text{4 pixels have been replaced with data bits} \\
&\text{in such way that two pixel value} \\
&\text{increases by 1 and other decreases by 1} \\
&\quad \vdots \\
&\ \text{OR} \\
&\text{16 pixels have been replaced by data bits} \\
&\text{in such way that for 8 pixels the value} \\
&\text{increases by 1 and other decreases by 1}\} \\
= 1 - \{&(1 - i/2)^{16} \\
&+(1 - i/2)^{14} * (i/2)^2 * 16C_2 * p' * (1 - p') * \frac{2!}{1! * 1!} \\
&+(1 - i/2)^{12} * (i/2)^4 * 16C_4 * p'^2 * (1 - p')^2 * \frac{4!}{2!2!} \\
&+(1 - i/2)^{10} * (i/2)^6 * 16C_6 * p'^3 * (1 - p')^3 * \frac{6!}{3!3!} \\
&\quad \vdots \\
&+(i/2)^{16} * 16C_{16} * p'^8 * (1 - p')^8 * \frac{8!}{4!4!}
\end{aligned} \tag{2}
$$

It is logically correct that $\eta_{ki}$ increases with $i$. Also it can be seen from equation 2 for $Pr$, that $\eta_{ki}$ increases with $i$ for $0 \leq i \leq 1$. This can be proved by differentiating equation 2 w.r.t. $i$ or can be empirically verified with ease. A close look at the equation reveals that $Pr$ depends upon $(p' * (1 - p'))^{(\text{Some positive integer power})}$. Given a Start image $S_k$, $p'$ is fixed. But which in turn depends upon $p$ and $k$. (Refer to Eq. 1). As $k$ increases to values of 1, $p'$ goes to $\frac{1}{2}$ irrespective of $p$. In general, adjacent pixels are very similar in natural images. So, $p$ is biased towards 0.35 or 0.65.

$$
\begin{aligned}
&\Rightarrow p' * (1 - p') \text{ increases} && \text{as } k \text{ increases,} \\
&\Rightarrow \qquad Pr \text{ decreases} && \text{as } k \text{ increases.} \\
&\Rightarrow \qquad \eta_{ki} \text{ decreases} && \text{as } k \text{ increases.}
\end{aligned}
$$

Thus, as $Pr$ decreases with increasing $k$, the number of wavelet coefficients of $S_k$ and $S_{ki}$ that are equal, increases, i.e. noise in $W(S_{ki})$ w.r.t $W(S_k)$ decreases.
$\Rightarrow \Gamma_W^{ki}$ increases as $k$ increases.

We have verified these experimentally as follows.

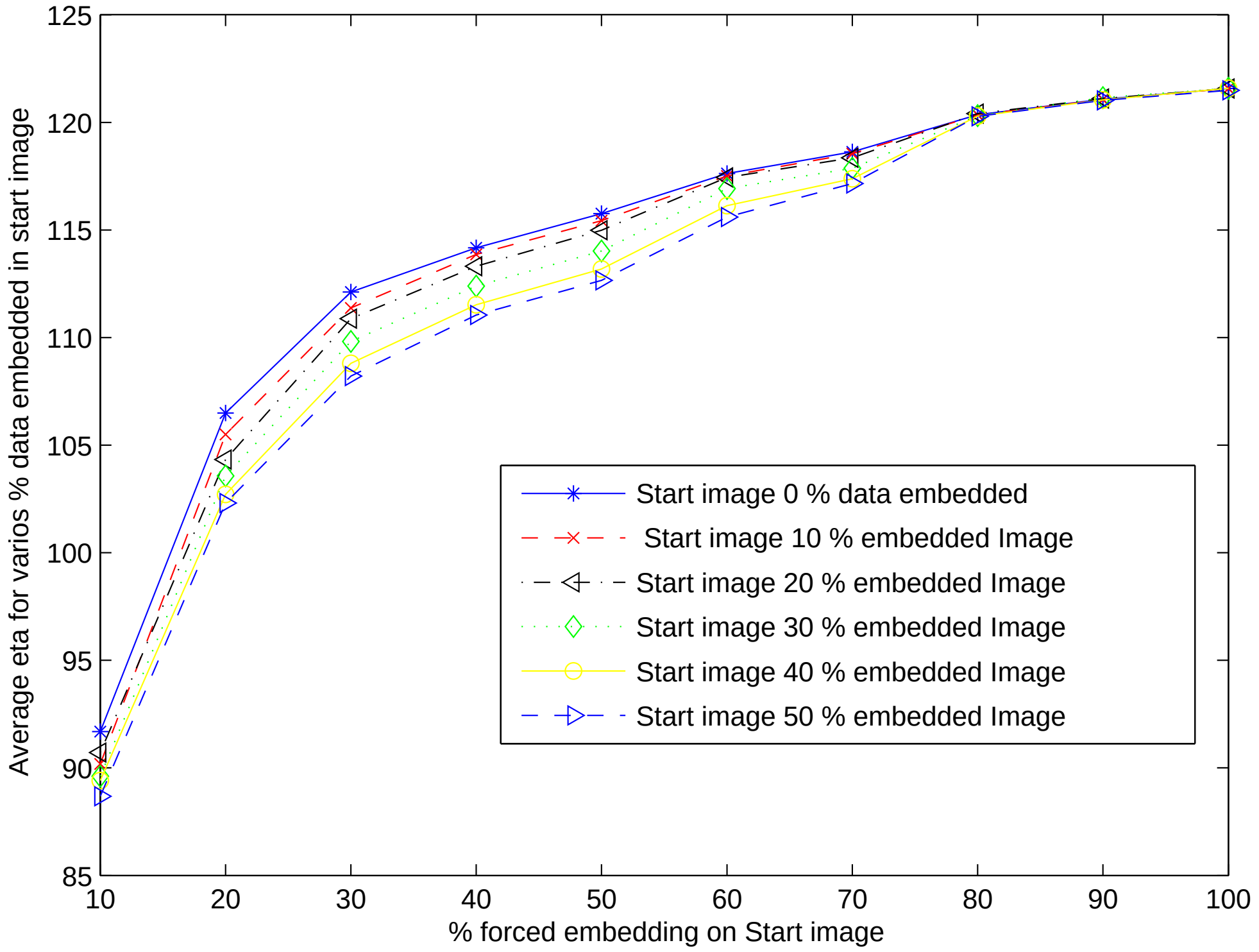

**Fig. 2.** *Graph of* $\eta_{ki}$ *vs 'i' for various 'k' Hide4PGP*

### 3.6 Results

We use the stego algorithm Hide4PGP in our experiments. In our experiments, we use $i = 10, \ldots, 100$. $k = 0, 10, 20, 30, 40, 50$. The plots of $\eta_{ki}$ vs. $i$ for various $k$ is as shown in Fig. 2.

For a particular forced embedding say $i$, it can be observed that $\eta_{ki}$ decreases as $k$ increases. Encouraged by this monotonic trend, we now look closely at the variations in measure $\eta$ at a fixed forced embedding of $i = 20\%$, with respect to $k$ on different start images. The results are shown in Fig. 3.

The continuous line shows the average value, $\eta_{k20}$ vs. $k$. The other curves show the $\eta$ values for the individual images. These also show the monotonic decreasing trend around the average value. We note that such trends are quite significant especially at low levels of 20% embedding. Thus, this serves as a first indicator for detecting approximately the amount of embedding (even at low levels) in any given image.

It is quite difficult to conduct a large number of data generation experiments under various parameter choices using a public domain tool as we do not get appropriate handles into the source code. Hence, in our lab we have built a tool called CSA-Tool for simulating the behavior of S-Tool. We have taken care to incorporate our own functions for encryption, randomized location generation and embedding analogous to the steps performed by S-Tools. The statistical characteristics of our tools would closely resemble those of S-Tools.
We performed similar experiments as detailed above using the CSA tool. Fig.4 and Fig.5 show the results. We note that the results are along the same trends as for Hide4PGP. However, the separations in Fig.4 are smaller than in Fig. 2 and fluctuations in Fig.5 are more than in

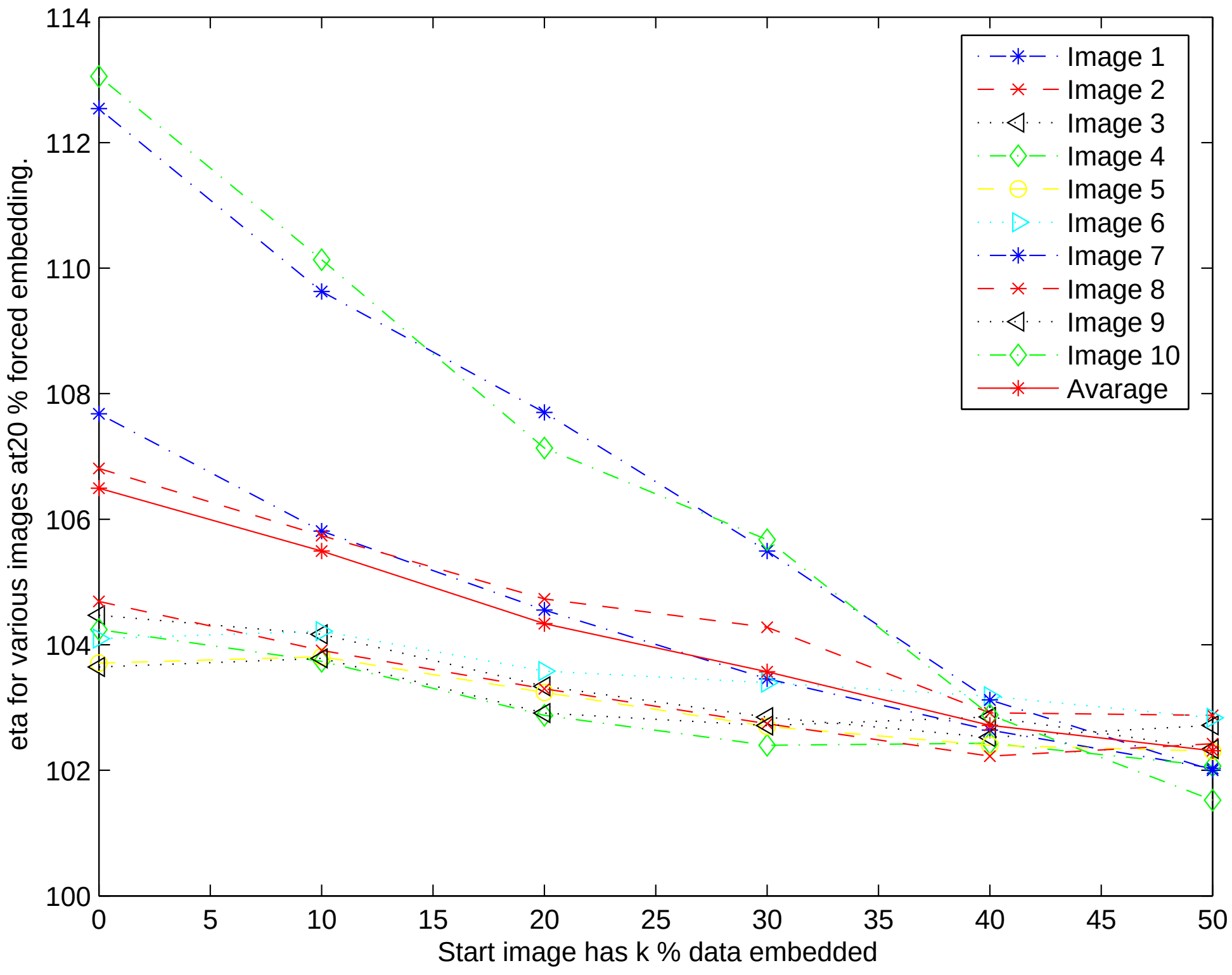

**Fig. 3.** *Graph of $\eta$ vs 'k' for various at fixed forced embedding 20% for various images Hide4PGP*

Fig. 3. A reason is that the CSA Tool (and S-Tools) employs more strong random generators for choosing the LSB for embedding than the tool Hide4PGP.

The plot, $\Gamma_W^{ki}$ vs. $i$ for various $k$ is as shown in Fig. 6. As per the theorem proved in Section 3.5, it can be observed that for a particular forced embedding say $i$, $\Gamma_W^{ki}$ increases as $k$ increases. The zoomed version of Fig. 6, for $i = 70$ is shown in Fig. 7. Encouraged by this monotonic trend, we now look closely at the variations in measure $\Gamma_W^{ki}$ at a fixed forced embedding of $i = 70\%$, with respect to $k$ on different start images. The results are shown in Fig. 3.6. The continuous line shows the average value, $\Gamma_W^{k70}$ vs. $k$. The other curves show the $\Gamma_W^{k70}$ vs. $k$ values for the individual images. These also show the monotonic decreasing trend around the average value.

## 4 Conclusion

We discussed two of our approaches towards analysis of stego images for detection of levels of embedding. Our approach of using wavelet coefficient perturbations holds promise. We plan to use this measure in addition to a statistical measures to arrive at finer detection in future.

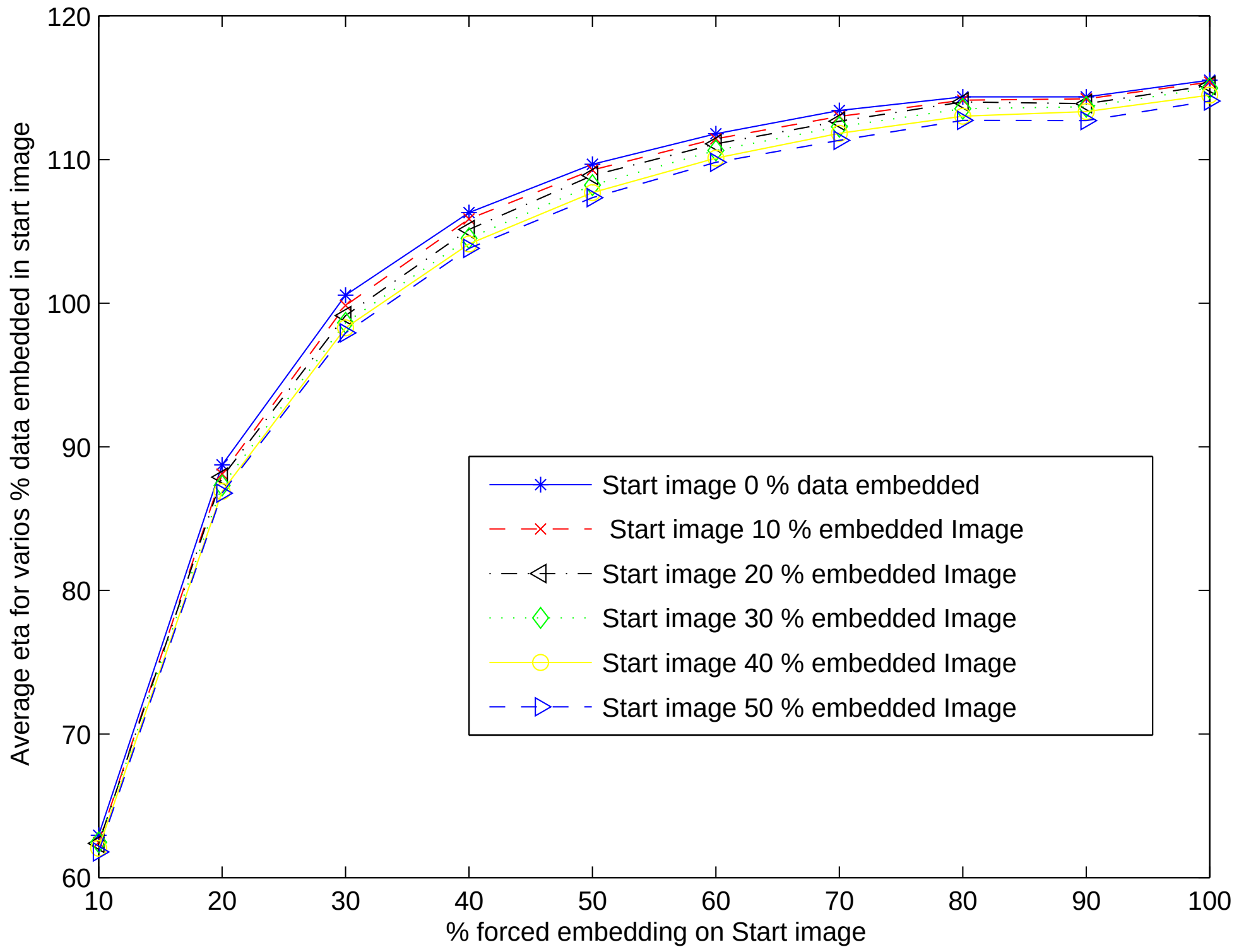

**Fig. 4.** *Graph of $\eta_{ki}$ vs. 'i' for various 'k' CSA Tool*

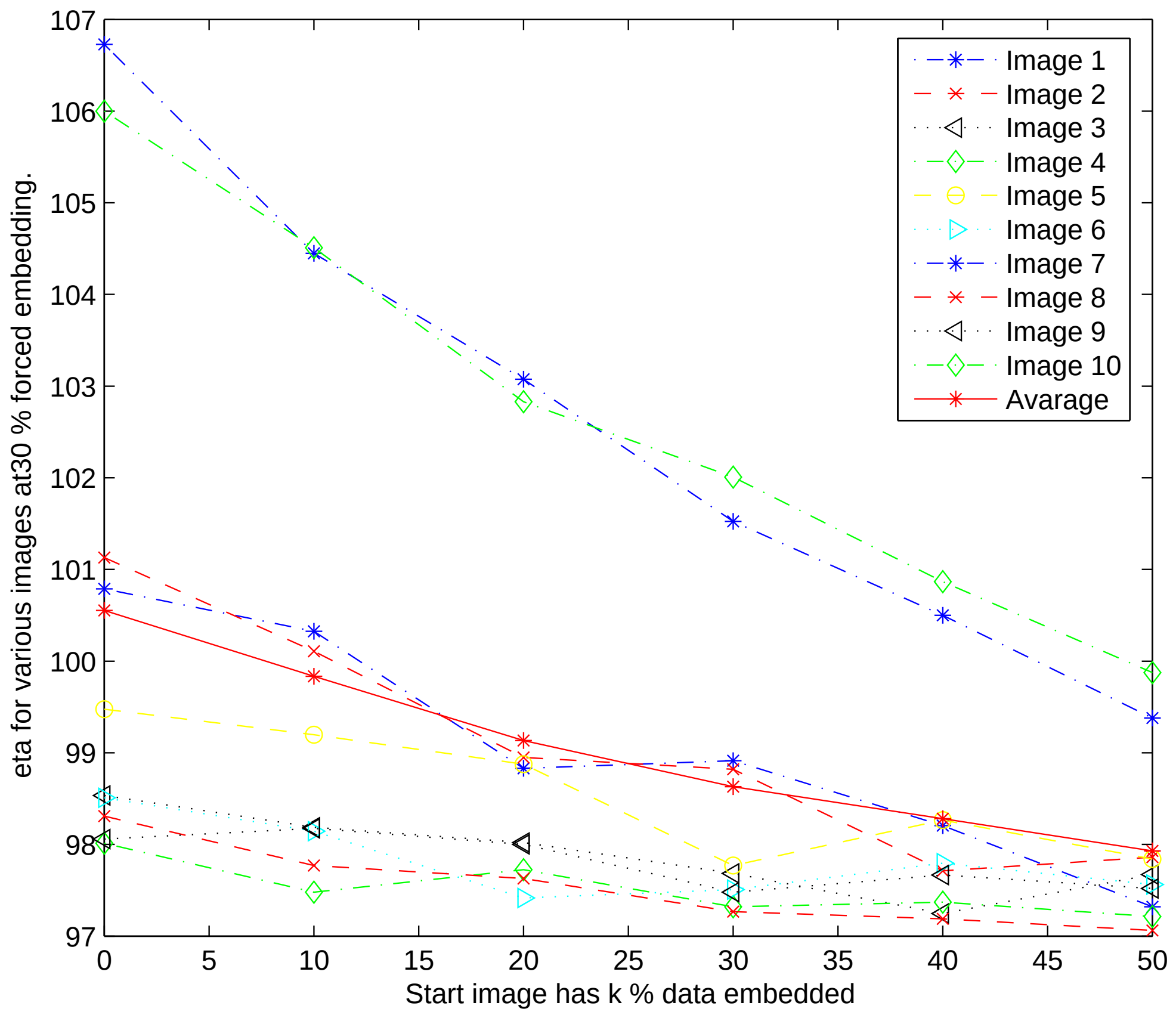

**Fig. 5.** *Graph of $\eta$ vs. 'k' for various at fixed forced embedding 30% for various images CSA Tool*

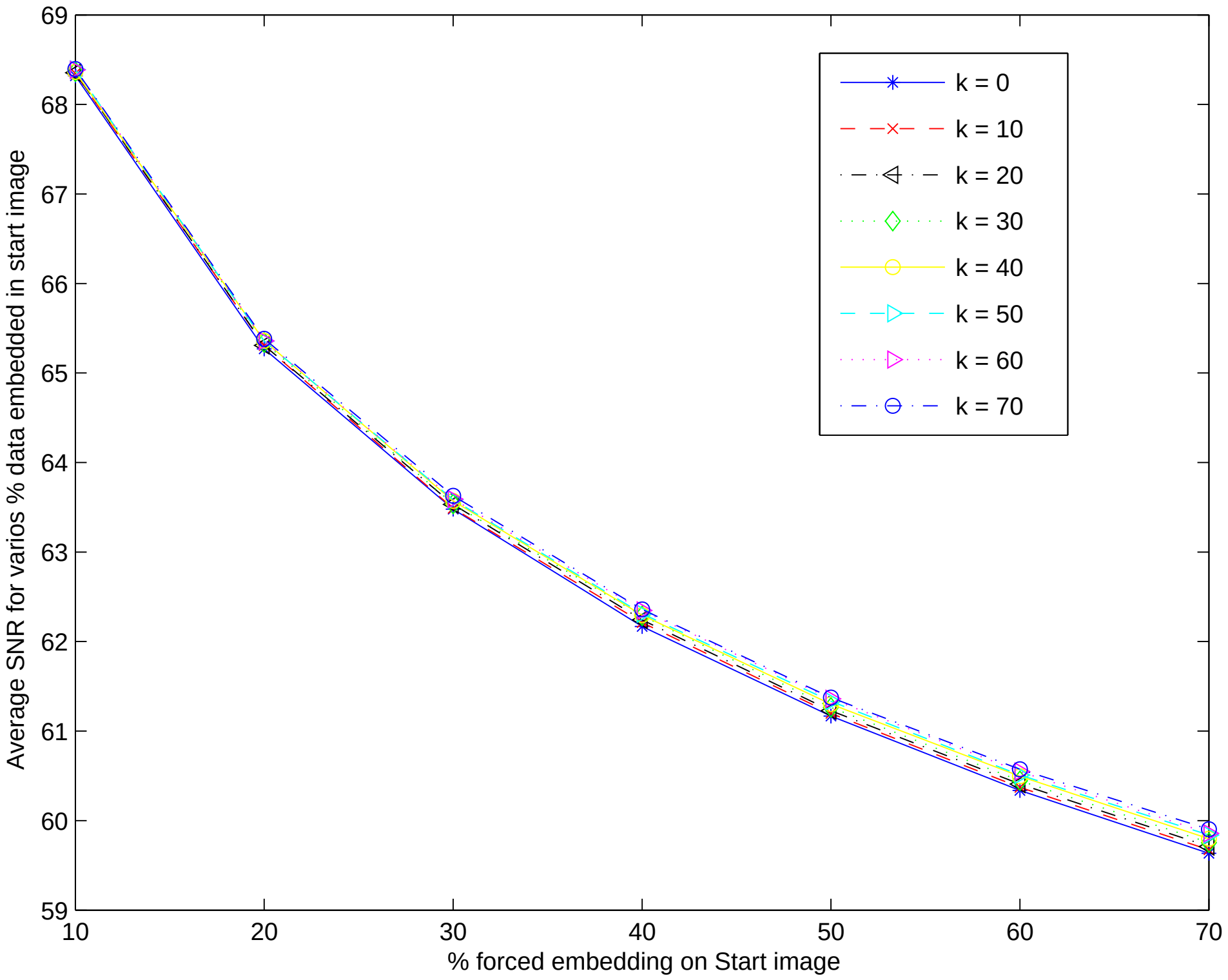

**Fig. 6.** *Graph of* $\Gamma_W^{ki}$ *vs. 'i' for various 'k' CSA Tool*

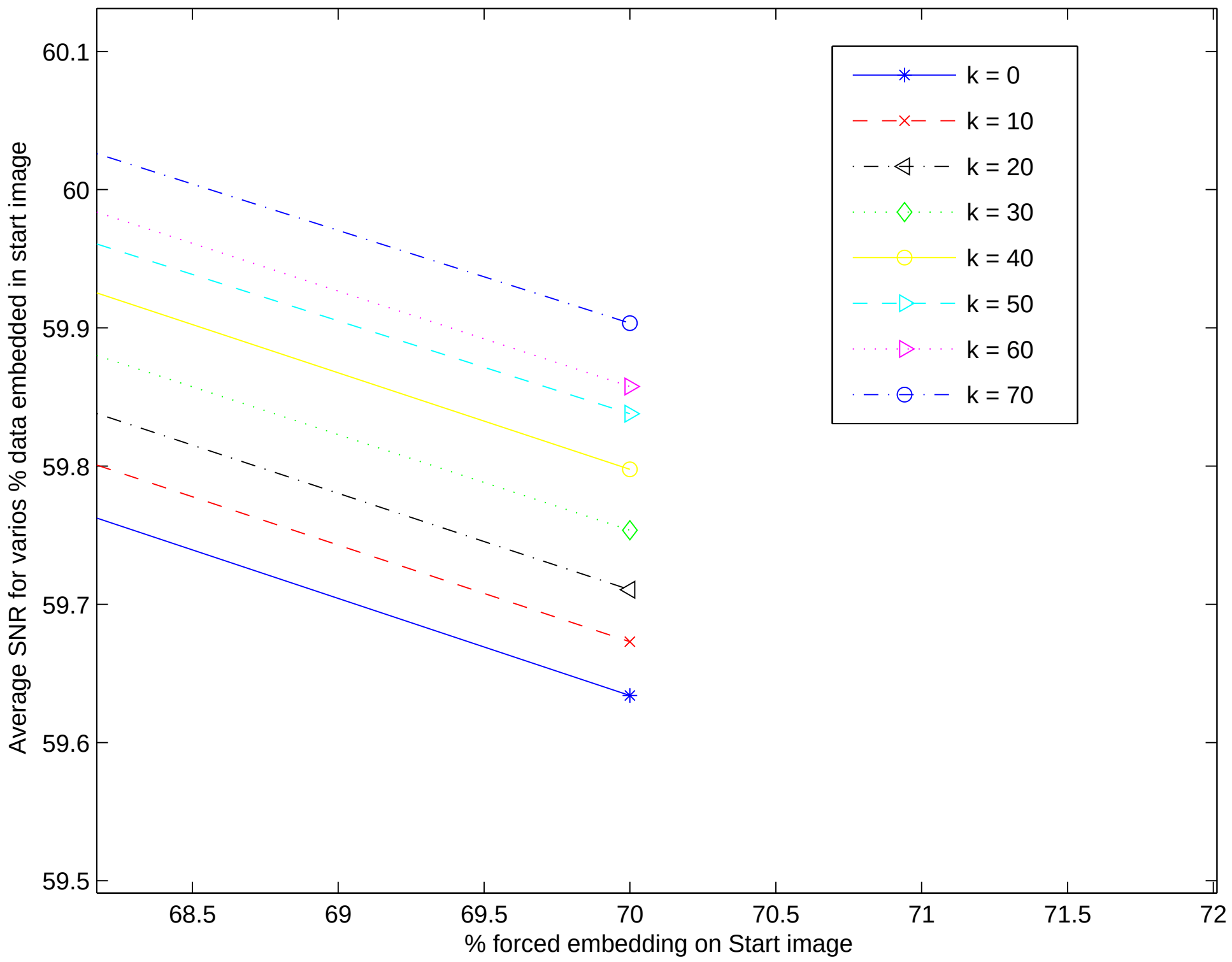

**Fig. 7.** *Graph of* $\Gamma_W^{ki}$ *vs. 'i' for various 'k' CSA Tool - zoomed version*

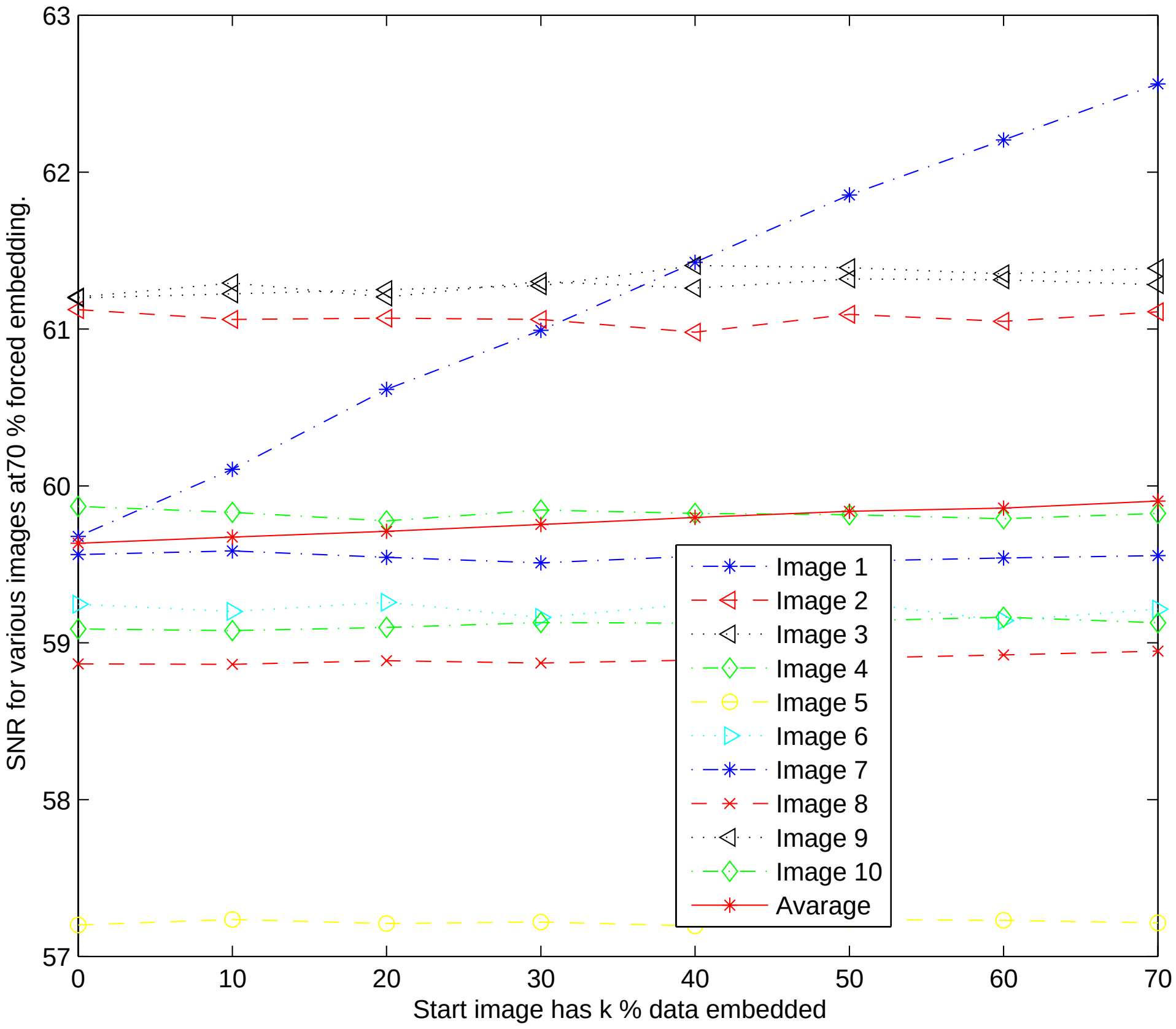

**Fig. 8.** *Graph of* $\Gamma_W^{ki}$ *vs. 'k' for various at fixed forced embedding 70% for various images CSA Tool*